\documentclass{sigkddExp}

\usepackage{times}  
\usepackage{helvet} 
\usepackage{courier}  
\usepackage[hyphens]{url}  
\usepackage{graphicx} 
\usepackage{caption} 
\usepackage{framed}
\usepackage{wrapfig}
\usepackage[utf8]{inputenc} 
\usepackage{makecell}
\usepackage[T1]{fontenc}    
\usepackage{hyperref}       
\usepackage{url}            
\usepackage{booktabs}       
\usepackage{amsfonts}       
\usepackage{nicefrac}       
\usepackage{microtype}      
\usepackage{subcaption}
\usepackage{amsmath}
\usepackage{comment}

\usepackage{times}
\usepackage{epsfig}
\usepackage{graphicx}
\usepackage{amsmath}
\usepackage{amssymb}
\usepackage[lined]{algorithm2e}
\usepackage{algorithmic}
\usepackage{chngcntr}

\DeclareMathOperator*{\argmin}{\bf argmin}
\newcommand{\vect}[1]{\mbox{\boldmath$#1$}}

\begin{document}
%

\title{Efficient Encrypted Inference on Ensembles of Decision Trees}
%

%


\author{
%
Kanthi Sarpatwar \\
       \affaddr{IBM Research AI}\\
       \affaddr{New York, USA}\\
       \texttt{sarpatwa@us.ibm.com}\\
       \and 
Karthik Nandakumar\\
       \affaddr{Mohamed Bin Zayed University of AI}\\
       \affaddr{Abu Dhabi, UAE}\\
       \texttt{karthik.nandakumar@mbzuai.ac.ae}
\and Nalini Ratha\\
       \affaddr{University at Buffalo, SUNY}\\
       \affaddr{New York, USA}\\
       \texttt{nalini.ratha@gmail.com}
 \and James Rayfield\\
       \affaddr{IBM Research AI}\\
       \affaddr{New York, USA}\\
 \texttt{jtray@us.ibm.com}\\
\and Karthikeyan Shanmugam\\ 
       \affaddr{IBM Research AI}\\
       \affaddr{New York, USA}\\
\texttt{karthikeyan.shanmugam2@ibm.com}\\
\and  Sharath Pankanti\\  Microsoft Research\\
\texttt{sharath.pankanti@gmail.com}\\
\and Roman Vaculin\\
       \affaddr{IBM Research AI}\\
       \affaddr{New York, USA}\\
 \texttt{vaculin@us.ibm.com}
}



\maketitle
\begin{abstract}
Data privacy concerns often prevent the use of cloud-based machine learning services for sensitive personal data. While homomorphic encryption (HE) offers a potential solution by enabling computations on encrypted data, the challenge is to obtain accurate machine learning models that work within the multiplicative depth constraints of a leveled HE scheme. Existing approaches for encrypted inference either make ad-hoc simplifications to a pre-trained model (e.g., replace hard comparisons in a decision tree with soft comparators) at the cost of accuracy or directly train a new depth-constrained model using the original training set. In this work, we propose a framework to transfer knowledge extracted by complex decision tree ensembles to shallow neural networks (referred to as DTNets) that are highly conducive to encrypted inference. Our approach minimizes the accuracy loss by searching for the best DTNet architecture that operates within the given depth constraints and training this DTNet using only synthetic data sampled from the training data distribution. Extensive experiments on real-world datasets demonstrate that these characteristics are critical in ensuring that  DTNet accuracy approaches that of the original tree ensemble. Our system is highly scalable and can perform efficient inference on batched encrypted (134 bits of security) data with amortized time in milliseconds. This is approximately three orders of magnitude faster than the standard approach of applying soft comparison at the internal nodes of the ensemble trees. 
\end{abstract}
\section{Introduction}

Machine Learning as a Service (MLaaS) is now a popular paradigm, where pre-trained models are hosted on a public cloud and inference is performed on a pay-per-query basis. However, the use of MLaaS is restricted in many application domains (e.g., financial services, healthcare) because the privacy (confidentiality) of the client data on which the inference is performed is of utmost concern. For example, consider a common setting involving a financial credit rating firm and a leasing agency. The credit rating firm owns a proprietary credit scoring ML model and offers the rating service on a public cloud. The leasing agency wishes to use that service to check its customers' financial worthiness. Without rigid privacy constraints, the leasing agency would be forced to send the raw (sensitive) financial data of its customers to the cloud. This would put the agency's customers at a significant risk of identity and financial theft, which in turn may result in considerable liabilities on the agency's end. Such privacy concerns are no longer fears of isolated individuals, but have become legal requirements due to privacy laws such as General Data Protection Regulation (GDPR) in Europe. 

Fully homomorphic encryption (FHE) can solve the above privacy conundrum by allowing certain types of computations on encrypted data without the need for decryption~\cite{acar2018survey}. An FHE scheme can be defined as: $H = (\mathcal{E}, \mathcal{D}, \lambda, \textsc{Eval})$, where $\mathcal{E}$ and $\mathcal{D}$ represent encryption and decryption operations, respectively, $\lambda$ is the security parameter, and $\textsc{Eval}$ is the evaluate function, which takes an arbitrary function $f$ and an encrypted input $\mathcal{E}(I)$ and returns the encrypted result $\textsc{Eval}(f, \mathcal{E}(I)) = \mathcal{E}(f(I))$. For ease of presentation, we ignore any mention of private and public keys, but it is understood that  $\mathcal{E}$ and $\textsc{Eval}$ require access to the public key and $\mathcal{D}$ needs access to the private key.

\begin{figure}[htpb]
\begin{center}
   \includegraphics[width=\linewidth]{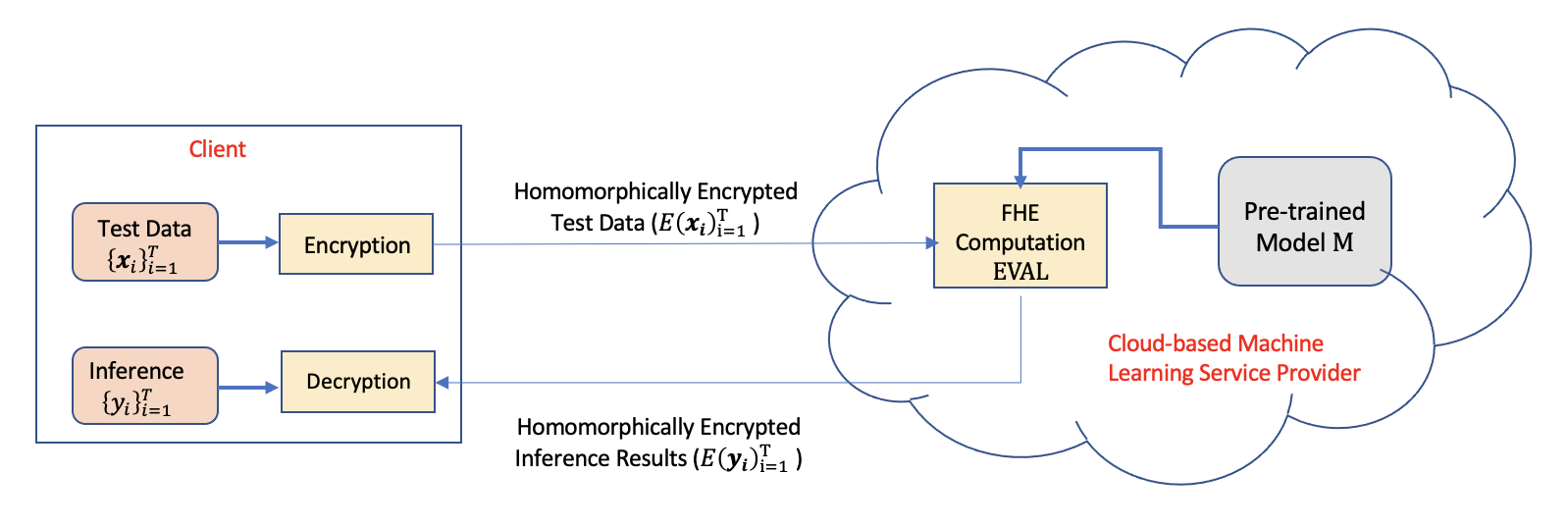}
\end{center}
   \caption{Scenario for \emph{non-interactive} privacy-preserving machine learning as a service (MLaaS) using fully homomorphic encryption (FHE).}
\label{fig:mlaas1}
\end{figure}

In this paper, we focus on FHE-based privacy-preserving inference scenario (see Figure \ref{fig:mlaas1}), which works as follows. The client encrypts its data $\vect{x}$ and sends the encrypted data $\mathcal{E}(\vect{x})$ to the cloud. The service provider, who holds a pre-trained model $\mathcal{M}_\theta$ ($\mathcal{M}$ and $\theta$ denote the model architecture and parameters, respectively), performs inference computation in the encrypted domain ($\mathcal{E}(y) = \textsc{Eval}(\mathcal{M}_\theta, \mathcal{E}(\vect{x}))$) and returns the encrypted result back to the client for decryption. The key point to note is that the inference service does not require the private key to perform its inference computation. Therefore, the service provider does not gain any knowledge about the input ($\vect{x}$) or the resulting output ($y$). Thus, complete client data privacy (confidentiality) is guaranteed. Furthermore, the above scenario is referred to as \emph{non-interactive} because it involves only a single round of communication between the client and the service provider. It must be emphasized that the depicted scenario is still vulnerable to other kinds of black-box attacks (e.g., substitute model training, membership inference attack, etc.). This happens because inference computations may leak sensitive information about the training data or model parameters \cite{PapernotPATE2018}.  Addressing such attacks is beyond the scope of this work, which focuses only on client data privacy.

The fundamental challenge in encrypted inference arises from the serious restrictions in terms of the nature of feasible computations and the efficiency of computations in the encrypted domain. Most FHE schemes are constructed based on a \emph{leveled} HE scheme, which supports computations only up to a preset level of multiplicative depth determined by its parameters \cite{brakerski2014leveled}. This is because the ``noise'' in the ciphertexts keeps growing with each HE computation and decryption will fail once the noise grows beyond a threshold. This depth constraint can be surmounted using bootstrapping \cite{gentry2009fully}, which ``refreshes'' the ciphertext and reduces its noise level (at the cost of relying on circular security). However, bootstrapping is a highly expensive and complex operation that must be avoided or used sparingly for the computations to be practically feasible. In ML context, the problem of limited depth is further exacerbated by the presence of non-linear computations (e.g., comparison of two numbers, activation functions), which require a larger depth because the underlying HE schemes often support only additive and multiplicative homomorphism. Due to these reasons, it is seldom possible to directly use any pre-trained ML model for encrypted inference. 

Recent works on encrypted inference have attempted to solve this problem in two ways. One approach is to train an unconstrained model and then simplify its architecture to reduce the multiplicative depth, without re-training the parameters \cite{gilad2016cryptonets,xiao2019mca}. The other alternative is to define a depth-constrained model architecture and directly train it from scratch using the available training data \cite{HesamifardEncryptedDeepNN2019,sanyal2018tapas}. While the former approach typically leads to significant loss of inference accuracy compared to the original model, the latter approach may not be possible in practice because the training data used to train the original model may no longer be available\footnote{In many practical ML applications, it is not possible to indefinitely retain the data used to train a ML model because of privacy regulations (see storage limitation principle in Article 5.1.(e) of GDPR). Consequently, when there is a need to transform the model $\mathcal{M}_{\theta}$ (that may have been trained sometime in the past) into its depth-constrained version, the original training data may not be available.}.

In this work, we focus on efficiently enabling (homomorphically) encrypted inference on decision tree based ensemble methods such as Gradient Boosting, AdaBoost, or XGBoost. This is because boosted decision trees \cite{chen2016xgboost} are still the standard choice for classification and regression tasks over tabular datasets \cite{popov2019neural}.

{\bf Challenges in Encrypted Decision Tree Inference:} Directly applying decision trees on encrypted data is difficult primarily due to their reliance on comparison operations. This problem becomes more pronounced in encryption schemes that do not rely of bit-wise operations such as CKKS~\cite{cheon2017homomorphic}, which directly operates on floating-point numbers. Consequently, performing encrypted inference even on an individual decision tree is computationally expensive in terms of both runtime and memory requirement. Since practical ensemble methods comprise hundreds of decision trees, the above problem gets further amplified.

One potential approach to address the above challenge is to replace the non-linear hard comparison operation in decision trees by a ``softer'' operation. Examples of such soft  operations include ``scaled-up'' \textsc{sigmoid} function (Eq. \ref{scaled_sigmoid}) or computing the power function in Eq. \ref{power_func} for a large positive $k$ \cite{cheon2019numerical}.

\begin{equation}\label{scaled_sigmoid}
\frac{1}{1+e^{-\alpha (a-b)}} \approx \begin{cases}
1 &{a > b}\\
0 &{a < b}
\end{cases}
\end{equation}

\begin{equation}\label{power_func}
\frac{a^k}{a^k + b^k} \approx \begin{cases}
1 &{a > b}\\
0 &{a < b}
\end{cases}
\end{equation}

The key problem with the scaled \textsc{sigmoid} approximation is that the scaling factor $\alpha$ depends on the absolute difference between $a$ and $b$, and Eq. \ref{scaled_sigmoid} works well when $\alpha \geq \frac{8}{|a-b|}$. In a decision tree, one needs to compare feature values and corresponding thresholds at different nodes of the tree. In order to adopt the scaled \textsc{sigmoid} function, one must choose $\alpha = \max_{n\in T} \frac{8}{|t_n - f_n|}$ where $t_n$ is the threshold value and $f_n$ is the feature value for each node $n$ of the decision tree $T$. Unfortunately, this may increase the value of $\alpha |a-b|$ to an arbitrarily large number and since HE computations can be applied only on numbers in a certain bounded range, we cannot use this approximate comparison method. Similarly, the accuracy of the power function approximation depends on the ratio between $a$ and $b$. When $a$ and $b$ are close, a large value of $k$ would be required. Furthermore, computing the function in Eq. \ref{power_func} in the encrypted space is extremely expensive because it involves a division operation and may need multiple refreshes (\emph{bootstrapping}) of the ciphertext. 

{\bf Our Contributions:} The above limitations of the soft comparison approach motivates us to address the problem at the training step by building models that are ``HE-friendly''. Hence, we propose a framework to learn a HE-friendly surrogate model that mimics the functionality of the original ensemble decision tree model. While training such a surrogate model has been employed before in the context of large neural network ensembles \cite{HintonKnowDist}, we are the first to use it in the context of encrypted inference for ensemble decision tree models. Towards this end, we make the following key contributions:

\begin{enumerate}

\item We are the first to propose a framework for transforming an ensemble of decision trees into a (multiplicative) depth-constrained shallow neural network model called DTNet. In fact, the proposed framework can transform any given pre-trained machine learning model $\mathcal{M}_{\theta}$ into its depth-constrained version $\widetilde{\mathcal{M}}_{\tilde{\theta}}$ without access to the original training data, while minimizing any accuracy degradation. 

\item We propose an algorithm to search for the best DTNet model that can be trained accurately even if the training set (used to train the original model) is not available.
    
\item We demonstrate the utility of the proposed framework to perform highly-efficient encrypted inference on complex ensembles (100s of base estimators) of decision trees. The accuracy and scalability of this approach has been demonstrated on several real-world public datasets.
        
\end{enumerate}

\section{Related Work}
\label{sec:relatedwork}
Privacy-preserving ML is a broad topic addressing a variety of threats \cite{al2019privacy}. In this work, we focus only the privacy threat arising from leakage of sensitive personal information when data is sent to an external service provider for inferencing. Existing solutions to this problem typically use some combination of cryptographic techniques such as garbled circuits, secure multiparty computation, and homomorphic encryption \cite{badawi2018alexnet,juvekar2018gazelle,liu2017oblivious,liu2020tifs,mohassel2017secureml,riazi2018chameleon}. We exclusively focus on the homomorphic encryption (HE) approach, which is non-interactive and guarantees provable privacy.

Encrypted inference has been made possible due to the advent of several FHE cryptosystems such as BFV \cite{brakerski2012fully,FV_FHE}, BGV \cite{brakerski2014leveled}, CKKS \cite{cheon2017homomorphic}, and TFHE \cite{chillotti2020tfhe}. Among these schemes, BFV and BGV support efficient vector operations over integers, the CKKS scheme allows ``approximate'' floating-point operations, and TFHE has faster bootstrapping for binary gates. Moreover, all these schemes support Single Instruction Multiple Data (SIMD) operations, which facilitates  parallelization of operations by packing different plaintext values into different ciphertext \emph{slots} (known as ciphertext packing).

Inference on encrypted data has been attempted for a variety of ML models ranging from logistic regression to deep neural networks. In a foundational work, Bost et al. \cite{bost2015ndss} identified and implemented a set of core operations over encrypted data (namely, comparison, argmax, and dot product), which can be combined to construct many classifiers. CryptoNets \cite{gilad2016cryptonets} used square function as the activation layer and a scaled mean pooling strategy to demonstrate encrypted inference over a convolutional neural network. This was improved upon in \cite{chabanne2017privacy}, where a low-degree polynomial approximating the ReLU function was used in combination with batch normalization. In both the above cases, a simplified version of a pre-trained network was used for inference. In contrast, Badawi et al. \cite{badawi2018alexnet} showed that training a HE-aware neural network from scratch can be beneficial. This was further validated in \cite{HesamifardEncryptedDeepNN2019}, where low-degree polynomial approximations for commonly used activation functions such as ReLU, Sigmoid, and Tanh were proposed and the networks were trained using these approximation polynomials. Sanyal et al. \cite{sanyal2018tapas} used a binary neural network and leveraged the efficient binary bootstrapping of TFHE scheme to achieve fast encrypted inference. 

While the challenge in encrypted inference over neural networks is finding a suitable approximation of the non-linear activation function, the corresponding challenge in decision trees is finding an efficient approximation of the comparison operation. In \cite{tueno2019arxiv}, the authors encoded the inputs to the comparator as bits and employed a BGV/TFHE cryptosystem to achieve secure decision tree evaluation. More recently, Xiao et al. \cite{xiao2019mca} proposed to replace the comparison function by a $24$-degree Chebyshev polynomial approximation of the sigmoid function to securely evaluate decision trees using FHE. To the best of our knowledge, there is no prior work that addresses the problem of homomorphic inference on an ensemble of decision trees.
\section{Proposed Approach}

The objective of this work is to construct a depth-constrained machine learning model ($\widetilde{\mathcal{M}}$) that is equivalent to (or closely mimics) the decision behavior of a given pre-trained model ($\mathcal{M}$). Note that the original ($\mathcal{M}$) and depth-constrained ($\widetilde{\mathcal{M}}$) learners need not belong to the same family of ML models. Here, depth refers to the multiplicative depth of the model, which is defined as follows.

{\bf Definition: (Multiplicative Depth of a Machine Learning Model)}
The multiplicative circuit depth of a machine learning model $\mathcal{M}$, denoted by $\Phi(\mathcal{M})$, is the maximum number of  multiplicative operations (gates) required along any of its circuit paths.

Let $\mathcal{M}:\mathcal{X} \rightarrow \mathcal{Y}$ be the pre-trained model, where $\mathcal{X}$ and $\mathcal{Y}$ denote the input and output spaces, respectively. Let $\mathcal{T} = \{(\vect{x}_i, y_i)\}_{i=1}^{T}$ be the training data used to learn the parameters $\theta$ of model $\mathcal{M}$, where $T$ is the number of training samples, and $p_{x}$ denote the training data distribution. Let $\mathcal{L}$ be a loss function that measures the prediction differences between $\mathcal{M}$ and its approximate version $\widetilde{\mathcal{M}}$. The problem of learning a depth-constrained model can be stated as:
\begin{equation}
    \argmin_{\widetilde{\mathcal{M}}_{\tilde{\theta}}}~ \mathbb{E}_{\vect{x} \sim p_x}~\mathcal{L}(\mathcal{M}_{\theta}(\vect{x}),\widetilde{\mathcal{M}}_{\tilde{\theta}}(\vect{x})), ~\widetilde{\mathcal{M}} \in \mathcal{H}_{\Omega} 
    \label{eqn:problemstatement}
\end{equation}
where $\mathcal{H}_{\Omega}$ is the depth-constrained hypothesis space, i.e., the set of all models whose multiplicative depth is less than or equal to $\Omega$. Since it is often impractical to directly estimate the expected loss in eq. (\ref{eqn:problemstatement}), we can replace it with the empirical risk. If the original training data $\mathcal{T}$ is available, the empirical risk minimization problem can be stated as:
\begin{equation}
    \argmin_{\widetilde{\mathcal{M}}_{\tilde{\theta}}}~ \sum_{i=1}^{T}\mathcal{L}(\mathcal{M}_{\theta}(\vect{x}_i),\widetilde{\mathcal{M}}_{\tilde{\theta}}(\vect{x}_i)), ~\widetilde{\mathcal{M}} \in \mathcal{H}_{\Omega} 
    \label{eqn:empiricalrisk}
\end{equation}

In this work, we assume that the original training data $\mathcal{T}$ is no longer available, which rules out the possibility of ``throwing'' out the given pre-trained model $\mathcal{M}_{\theta}$ and retraining a depth-constrained model from scratch. We propose to solve this problem using the knowledge distillation framework. \emph{Knowledge distillation} refers to the process of learning a student model that closely mimics the predictions of a teacher model \cite{HintonKnowDist}, which is typically achieved by allowing the teacher to weakly supervise the learning process of the student. Specifically, given a transfer set of input data samples (without true labels), the predictions made by the teacher model on this transfer set are used as the proxy labels to train the student model instead of the true labels. Algorithm \ref{alg:distillation} summarizes the main steps of our proposed approach for learning $\widetilde{\mathcal{M}}_{\tilde{\theta}}$ from $\mathcal{M}_{\theta}$.

\begin{algorithm}
    \caption{Depth-constrained Knowledge Distillation}
    \label{alg:distillation}
    \begin{algorithmic}
    \STATE \textbf{Inputs}: Pre-trained model $\mathcal{M}_{\theta}$, input data distribution $p_x$, multiplicative depth budget $\Omega$, distillation loss function $\mathcal{L}$, transfer set size $S$, validation set size $V$
    \STATE \textbf{Outputs}: Depth-constrained model $\widetilde{\mathcal{M}}_{\tilde{\theta}}$ that closely approximates $\mathcal{M}_{\theta}$ and its validation score $\nu$
    \newline 
    \STATE Step 1: Generate transfer set $\mathcal{S} = \{\tilde{\vect{x}}_i\}_{i=1}^{S}$ and validation set $\mathcal{V} = \{\bar{\vect{x}}_i\}_{i=1}^{V}$, where $\tilde{\vect{x}}_i, \bar{\vect{x}}_i \sim p_x$.
    \STATE Step 2: Select a model architecture $\widetilde{\mathcal{M}}$ from the hypothesis space $\mathcal{H}_{\Omega}$.
    \STATE Step 3: Learn $\tilde{\theta}$ as $\argmin_{\tilde{\theta}}~ \sum_{i=1}^{S}\mathcal{L}(\mathcal{M}_{\theta}(\tilde{\vect{x}}_i),\widetilde{\mathcal{M}}_{\tilde{\theta}}(\tilde{\vect{x}}_i))$.
     \STATE Step 4: Compute the validation score $\nu$ as the accuracy of the candidate model $\widetilde{\mathcal{M}}_{\tilde{\theta}}$ on the validation set $\mathcal{V}^{*} = \{\bar{\vect{x}}_i,\mathcal{M}_{\theta}(\bar{\vect{x}}_i)\}_{i=1}^{V}$.
    \end{algorithmic}
\end{algorithm}

Though knowledge distillation for model compression is a well-studied problem in the literature, it has never been applied to the field of designing models suitable for encrypted inference. Most existing works on model compression focus on making deeper neural networks shallow. We are the first to demonstrate that decision tree ensembles can be a good choice for teacher model especially for tabular data. Furthermore, the nature of the student model required for this application is very different. The goal is to obtain a network with low multiplicative depth and not simply any shallow network. A CNN with one convolution+ReLU+maxpool layer followed by a fully connected layer will be considered shallow by modern standards, but is almost impossible to implement in the encrypted domain (with reasonable speed) due to the high multiplicative depth of the max operation. This introduces an interesting additional complexity in the neural architecture search.

{\bf Transfer Set Generation:} The first question in knowledge distillation is how to obtain the transfer set of unlabeled input samples required for the distillation. In \cite{HintonKnowDist}, the original training data (minus the output labels) was used as the transfer set. The alternative is to employ synthetically generated \emph{pseudo training samples} that can be generated at the time of distillation. Since the teacher model has been trained only on the manifold $p_x$ of the training data, it's knowledge does not cover the whole input space. Hence, it has been argued that some knowledge of the input data manifold is required for synthetic transfer set generation \cite{liu2018teacher,lopes2017data}. More recently, attempts have also been made to perform knowledge distillation without access to any training data \cite{nayak2019zero,YooKEGNET}. However, these methods assume the teacher model is a deep neural network with a final softmax layer capable of producing the label distribution as the output. In this work, we assume that the service provider retains minimal information about the training data distribution $p_x$. Hence, we use a modified version of the MUNGE algorithm proposed in \cite{Bucilua_kdd_06}, where a non-parametric density estimate of $p_x(\vect{x})$ is obtained from a small subset of training data (without labels) and is used to generate a large unlabeled set that can be used for distillation.

{\bf Depth-constrained Student Network Architecture Search:}
The second question in knowledge distillation is the appropriate choice of the student network architecture. To the best of our knowledge, there has been no work in the literature that considers the problem of designing a student network with a limited multiplicative depth. The depth constraints complicate the search for the optimal student model in the following way. Suppose that a multi-layer perceptron with a sigmoid activation function is chosen as the student model. Given a limited multiplicative depth (say $\Omega$), it is not clear if the available depth should be earmarked for approximating the sigmoid function more accurately (say, using a higher-order polynomial) or adding more hidden layers to the network.

The classical work of Cybenko \cite{cybenko_89} showed that a fully-connected sigmoid neural network with a single hidden layer can universally approximate any continuous univariate function. Leshno et al. \cite{Leshno_91} extended this proof to any non-polynomial function as an activation function. Since the primary goal of our work is to approximate the decision function learned by the teacher model, we believe that a neural network is an appropriate choice for the student model. However, due to the limitations of FHE, we are restricted to polynomial activation functions. To compensate for this limitation, we add more hidden layers to the network to increase its modeling capacity. To determine the best student model architecture, we repeat Steps 2 through 4 of Alg. \ref{alg:distillation} for many choices of $\widetilde{\mathcal{M}} \in \mathcal{H}_{\Omega}$ and choose the student model with the highest validation score.

\subsection{Application to Homomorphic Inference of Decision Tree Ensembles} \label{sec:dtnets}

As a specific application of the proposed framework, we demonstrate how to perform encrypted inference with respect to large ensembles of decision tree classifiers. Unlike recent efforts that transfer knowledge from a complex model (neural networks and boosted trees) into a single decision tree to leverage the interpretable nature of decision trees \cite{bastani2017interpreting,frosst2017distilling}, our work is aimed at distilling the knowledge from a tree ensemble to a depth-constrained neural network (referred to as a {\em DTNet}) for facilitating inference on (homomorphically) encrypted data.

\begin{figure*}[h]
\begin{subfigure}[b]{0.386\linewidth}
    \includegraphics[width=\linewidth]{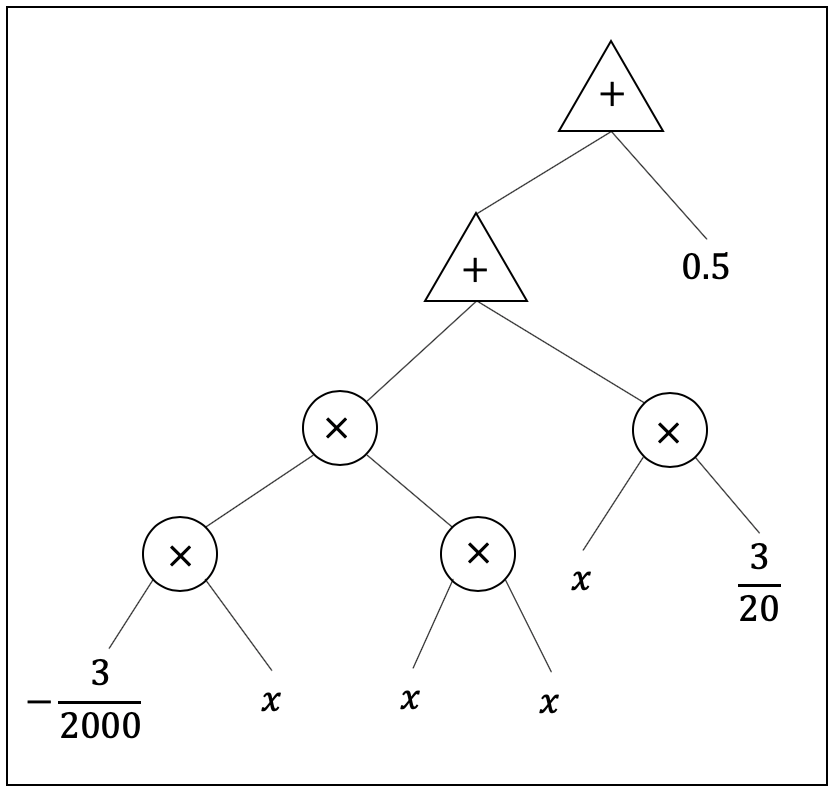}
        \caption{$\Phi(\textsc{approxSigmoid})=2$.}
    \label{fig:approxSigmoid}
\end{subfigure}
\begin{subfigure}[b]{0.515\linewidth}
    \includegraphics[width=\linewidth]{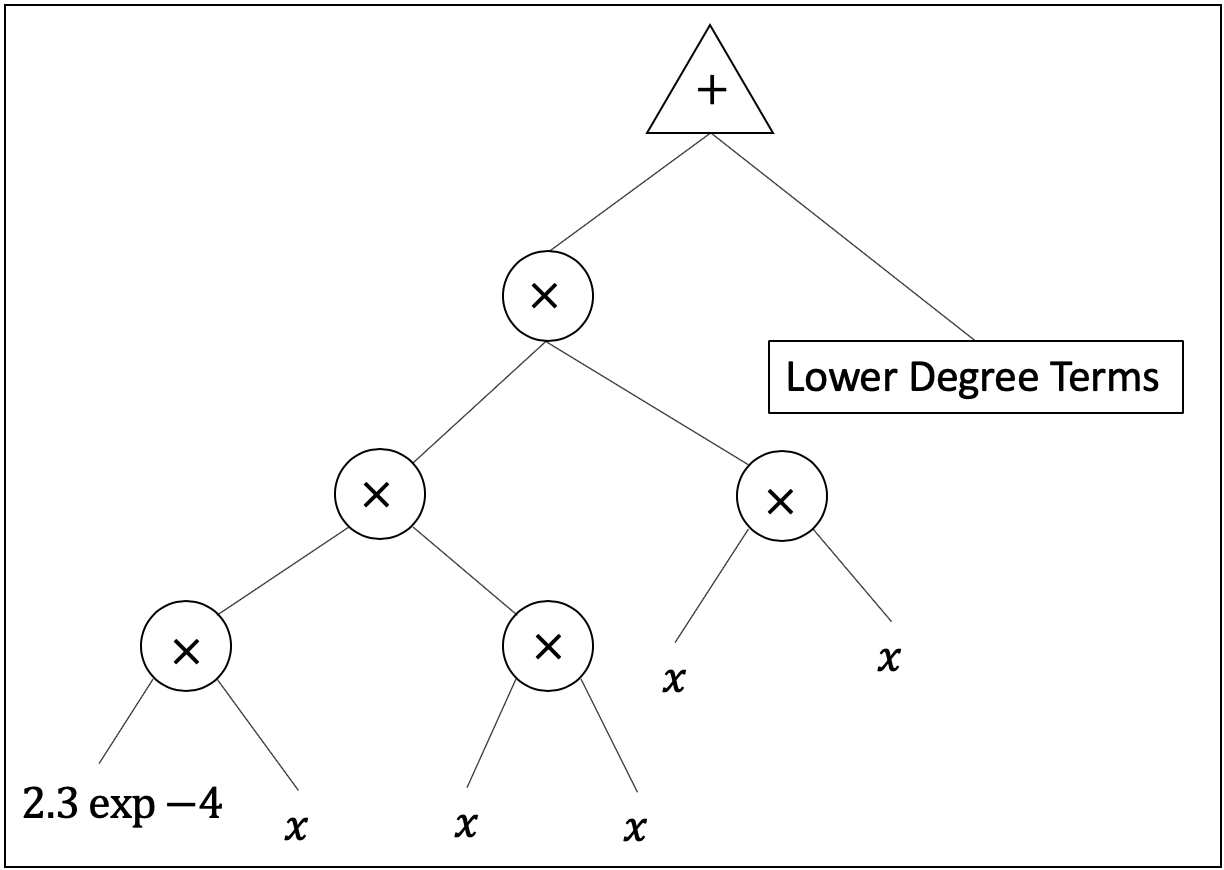}
        \caption{$\Phi(\textsc{approxRelu6})=3$ .}
    \label{fig:approxRelu}
\end{subfigure}
\caption{Computations of Multiplicative Circuit Depth ($\mathcal{M}(.)$) of Different Activation Functions\label{fig:circuit-depth}}

 \end{figure*}

We now describe the template structure of the DTNet neural network. The {\em input layer} comprises of $d$ nodes for input data with $d$ features. The {\em hidden layers} are dense (fully connected) layers with a specified number of nodes (\emph{neurons}) and a specified polynomial activation function. The list of polynomial activation functions that we have explored are shown in Table~\ref{table:activations} and the circuit depth computation for these functions are illustrated in Figure \ref{fig:circuit-depth}. The {\em output layer} is a softmax layer with $c$ nodes, where $c$ is the number of classes in the classification task. Note that when performing the inference, we can ignore the final softmax computation because it is a monotonic function that does not affect the relative class ordering. 

\begin{table*}[ht]
    \centering
    \begin{tabular}{|c|c|c|}
        \hline
        \thead{\bf Name} & \thead{\bf Description} & \thead{\bf  Multiplicative\\ \bf Circuit depth}\\
        \hline
         {\sc approxSigmoid} & $(-3.0/2000)x^3 + (3/20)x + 0.5$ & $2$\\
         \hline 
         {\sc approxRelu} & $0.47 + 0.50x+ 0.09x^2$ & $2$\\
         \hline
         {\sc approxRelu3} & $0.47 + 0.50x + 0.09x^2 - 1.7\exp(-10)x^3$ & $2$\\
         \hline
                  {\sc swish3} & $0.24 + 0.50x+ 0.10x^2 - 1.2\exp(-10)x^3$ &$2$\\
         \hline
        {\sc approxRelu6} & \makecell{$0.21 + 0.50x+ 0.23x^2+ 7.6\exp(-8)x^3 -$\\$ 1.1\exp(-2)x^4 - 3.5\exp(-9)x^5+ 2.3\exp(-4)x^6$} & $3$\\
         \hline
    \end{tabular}
    \caption{Descriptions of Various Polynomial Activation Functions Used and their Corresponding Circuit depths}
    \label{table:activations}
\end{table*}

An important virtue of DTNets is their low computational depth, which allows us to perform efficient and accurate homomorphic inference on these models. In fact, the inference predictions on the DTNets in the encrypted space are virtually identical to those in the unencrypted space. Further, as we demonstrate in the experimental section, the DTNets can approximate inference on an ensemble of decision trees quite closely even when the amount of training data available is very low. Hence, DTNets can be used to effectively approximate homomorphic inference on the tree ensemble.
\section{Empirical Results}
\label{sec:emp}
We now demonstrate the efficacy of our depth-constrained knowledge distillation approach for performing FHE inference on complex ensembles of decision trees. Our experimental analysis involves two dimensions: {\em homomorphic inference accuracy}, i.e., the accuracy of DTNets in comparison to that of the target ensemble, and {\em efficiency}, i.e., the time required to perform FHE inference on the DTNets. 

\subsection{Datasets}
\label{sec:datasets}
We use four datasets taken from public repositories (UCI~\cite{Dua:2019} and Kaggle~\cite{Kaggle}): Adult Consensus, Allstate Purchase Prediction Challenge~\cite{Kaggle}, Bank Marketing, and Magic Gamma Telescope dataset. These datasets offer a good mix of attribute types (categorical vs continuous) and their sizes vary between $20K$ to $600K$. While the Allstate dataset is a ternary classification problem, the remaining are binary classification problems. A brief summary of these datasets is shown in Table \ref{tab:my_label}. A detailed description of the datasets and preprocessing follows:

{\em Adult Consensus Dataset (ADULT):} This is a publicly available dataset comprising of $~32k$ points. We perform some standard preprocessing steps to encode the categorical various as integers (Scikit-Learn's LabelEncoder) and the continuous variables are scaled to lie between $[0,1]$.

{\em AllState Insurance Dataset (ALLSTATE):} This dataset is from a Kaggle prediction challenge. It contains $~665k$ records of insurance data belonging to Allstate customers. The objective we consider is that of predicting one of the products (named `A') which has three possible labels $0,1,2$. We drop a few columns that are irrelevant to the prediction task such as product types $A$-$G$, record\_type,  location and customer\_ID. Further, we impute missing values of continuous features by their respective means. Further, the continuous variables are scaled to lie in $[0,1]$

{\em Bank Management Dataset (BANK):} This dataset is again a UCI dataset comprising of $~45k$ records pertaining to data marketing campaigns conducted by a Portuguese financial institution. The objective is to predict whether a bank deposit term would be accepted (`yes') or not (`no'). The preprocessing steps are similar to the adult dataset. 

{\em MAGIC Gamma Telescope Dataset:}
This UCI dataset has $~19k$ points pertaining to registration of high energy gamma particles  in a ground-based atmospheric Cherenkov gamma telescope using the imaging technique. The preprocessing steps are the same as Adult and Bank datasets.

\begin{table*}[ht]
    \centering
    \begin{tabular}{|c
    |c|c|c|c|}
    \hline
        \thead{\bf Name} & \thead{\bf Dataset Size} & \thead{\bf Dimensions} & \thead{\bf Attribute Characteristics} & \thead{\bf Area} \\
        \hline
         ADULT & 48842 & 14 &Categorical, Continuous &Social\\
         \hline
         ALLSTATE & 665250 & 25 &Categorical, Continuous & Insurance\\
         \hline
         BANK Marketing & 45211 & 17 &Categorical, Continuous & Finance\\
         \hline
         MAGIC Telescope & 19020 & 11& Continuous &Science\\
         \hline
    \end{tabular}
    \caption{Description of Datasets}
    \label{tab:my_label}
\end{table*}

\subsection{Homomorphic Inference Accuracy} 

{\bf Obtaining the Target Ensemble:} For each experiment, we are given a pre-trained target model, which is a decision tree ensemble trained on the complete training dataset $\mathcal{T}$. The target ensemble is trained using the standard scikit-learn~\cite{scikit-learn} {\sc AdaBoost} classifier, with a depth $5$ decision tree classifier as a base estimator. We use a grid search to tune the hyperparameters of the model over the search space $\{\textit{learning rate}: [0.01,0.05,0.1,0.3,1], \textit{ number of estimators}: [50,100]\text{ }\}$. The best possible hyperparameters are chosen via a 4-fold cross-validation on the above hyperparameter space.

{\bf Obtaining the DTNet:} For the chosen ensemble, we obtain the DTNet using Alg. \ref{alg:distillation}. The size of the synthetically generated transfer set ($S$) is restricted to roughly $200K$-$300K$. We assume that a small subset of the original training data ($\widehat{\mathcal{T}}$), without true labels, is made available for obtaining the non-parametric estimates of $p_x$ required by the MUNGE algorithm \cite{Bucilua_kdd_06}. We vary the size of $\widehat{\mathcal{T}}$ to be in the order of $10^3$, $10^2$, and $10^1$ data points to measure its impact on the accuracy. While the validation set $\mathcal{V}$ could also be generated synthetically, for the final evaluation we assume that the same fraction of unlabeled training data (used for estimating $p_x$) is available as the validation set.

We use the standard TensorFlow library's Keras framework~\cite{tensorflow2015-whitepaper} to train the neural networks. For each architectural configuration, we train the corresponding neural network for $500$ epochs against a {\em sparse\_categorical\_crossentropy} loss function using an ADAM optimizer with a learning rate of $0.01$. For regularization, after each hidden layer, we add a dropout layer with a dropout rate of $0.02$. Finally, the best DTNet is chosen and used to perform encrypted  inference.

{\bf Architectural Search Space:} We now describe the architectural search space for the best DTNet model. We limit the number of layers in the DTNet to a maximum of $4$, including an input layer, output layer and two hidden layers. We use the DTNet template (see Section \ref{sec:dtnets}) to construct several specific neural networks skeletons with the following nodes configurations. {\em One Hidden Layer NN Configurations:}
$\mathcal{C}_1(\widetilde{\mathcal{M}}) := \{HL1:8,~HL1:16,~HL1:32,~HL1:64\}$. {\em Two Hidden Layer NN Configurations:}
$\mathcal{C}_2(\widetilde{\mathcal{M}}):=\{HL1|HL2:8|8,~HL1|HL2:16|16,~ HL1|HL2:32|32,~ HL1|HL2:64|64\}$. The multiplicative depth of a DTNet with one hidden layer and an activation function of depth $\Phi(A)$ is given by $(\Phi(A) + 2)$. Similarly, a DTNet with two layers, whose corresponding function depths are $\Phi(A_1)$ and $\Phi(A_2)$, is given by $(\Phi(A_1)+\Phi(A_2) + 3)$. It is important to note that even for the same skeleton of DTNet, different activation functions can lead to different circuit depth. Further, for each layer, one can apply one of the several activation functions listed in the Table~\ref{table:activations}. 

\begin{figure}[ht]
\centering
    \includegraphics[width=\linewidth]{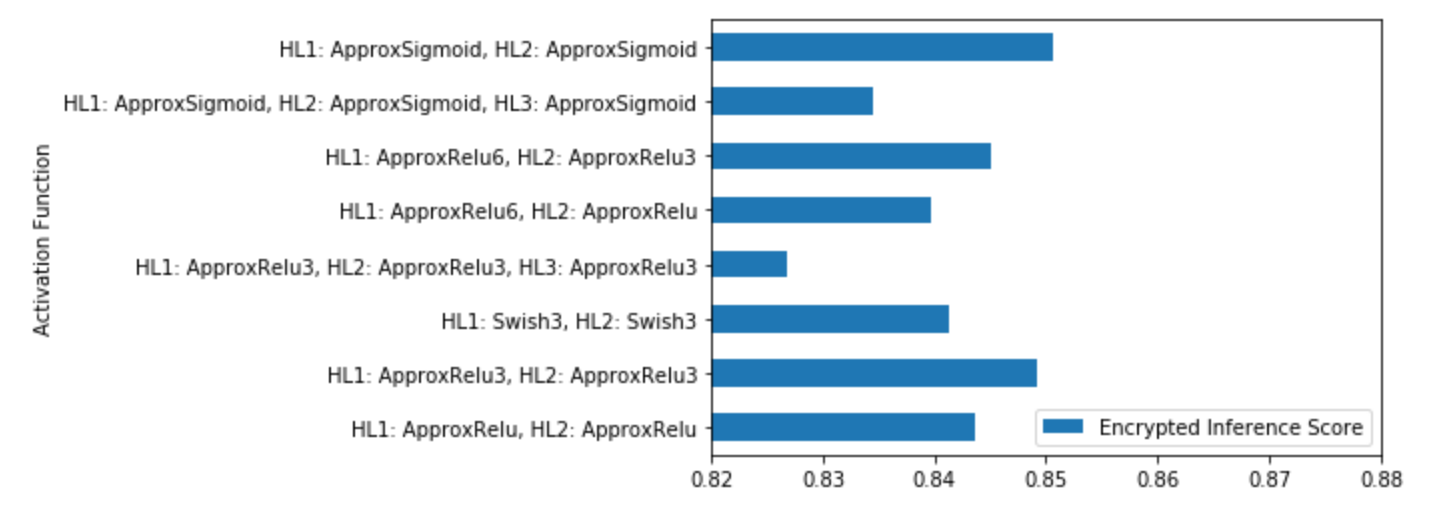}
        \caption{Impact of different activation functions on the DTNet accuracy using the ADULT dataset.}
    \label{fig:multiple-activations1}
\end{figure}

{\bf Architectural Search Results:} First, we study the impact of activation functions on the overall accuracy. For this experiment, we use the ADULT dataset and fix the network skeleton to two hidden layers, each containing $64$ nodes. We tried several configurations of activation functions listed in Table~\ref{table:activations} and compute the encrypted inference accuracy score. The results in  
Figure~\ref{fig:multiple-activations1} show that {\sc ApproxSigmoid} activation function outperforms all other activation functions. Hence, we use {\sc ApproxSigmoid} as the activation function in all the subsequent experiments.

\begin{figure}[ht]
\centering
    \includegraphics[width=\linewidth]{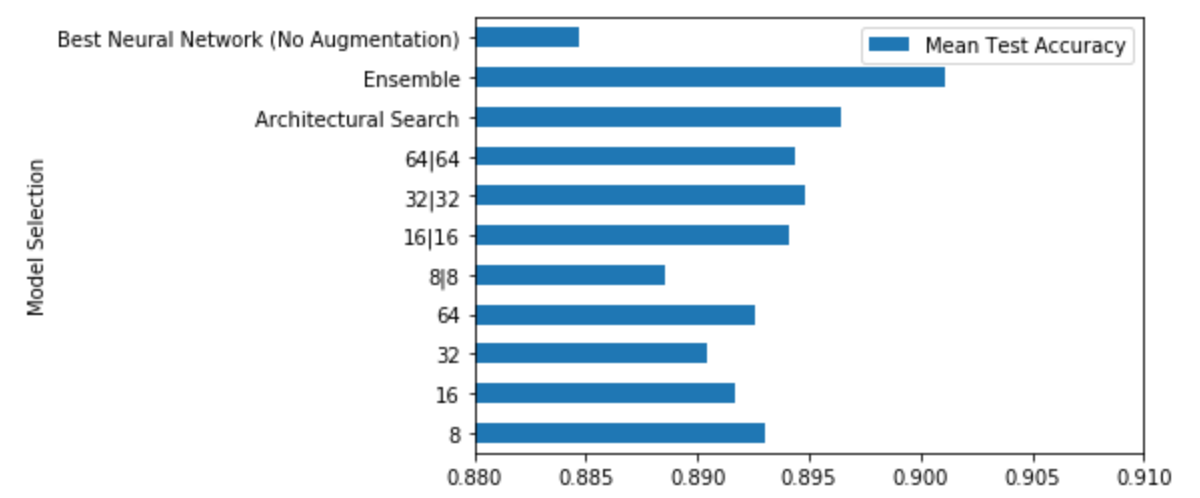}
        \caption{Impact of architectural search on the DTNet accuracy using the Bank dataset. Architectural search yields better accuracy compared to any {\em individual} architecture.}
    \label{fig:architecture1}
\end{figure}

Next, we attempt to determine the best neural network architecture in terms of number of hidden layers and number of neurons in each hidden layer. For this experiment, we chose the Bank dataset. To choose the best DTNet, we perform an architectural search over all possible configurations mentioned earlier in the search space discussion. Figure~\ref{fig:architecture1} shows the mean test accuracy of different configurations (e.g., $8$ represents one hidden layer with $8$ nodes and $8|8$ represents configuration with two hidden layers with $8$ nodes each) along with the mean test accuracy of the best model obtained by the architectural search. Clearly, the architectural search yields a better model compared to any individual architecture. 

{\bf Baseline:} As emphasized earlier, there is no prior work that satisfactorily addresses the problem of homomorphic inference on an ensemble of decision trees. So, as a baseline for comparing accuracy, we use the best possible depth constrained neural network we can train on the available training data ($\widehat{\mathcal{T}}$ with the true labels). This {\em simulates} the case where one ``throws away'' the ensembles and trains a depth-constrained model directly from the available training data. The implementation details of this baseline network are the same as those for DTNet training. The key difference is that the DTNet uses only the synthetic transfer set (generated from a small subset of unlabeled samples), while the baseline network uses the same subset of samples along with true labels. 

\begin{figure*}[ht]
\centering
\begin{subfigure}[b]{0.35\linewidth}
    \includegraphics[width=\linewidth]{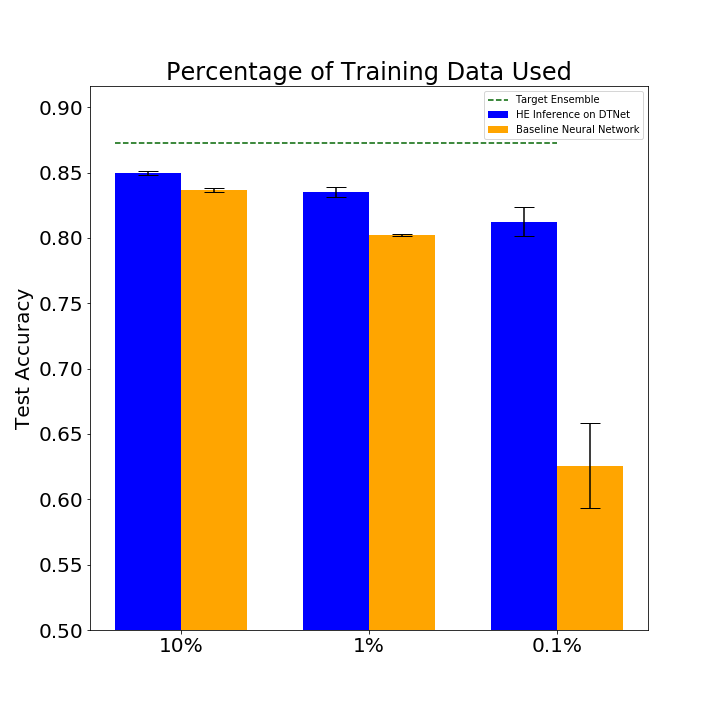}
    \caption{Adult Dataset}
\end{subfigure}
\begin{subfigure}[b]{0.35\linewidth}
    \includegraphics[width=\linewidth]{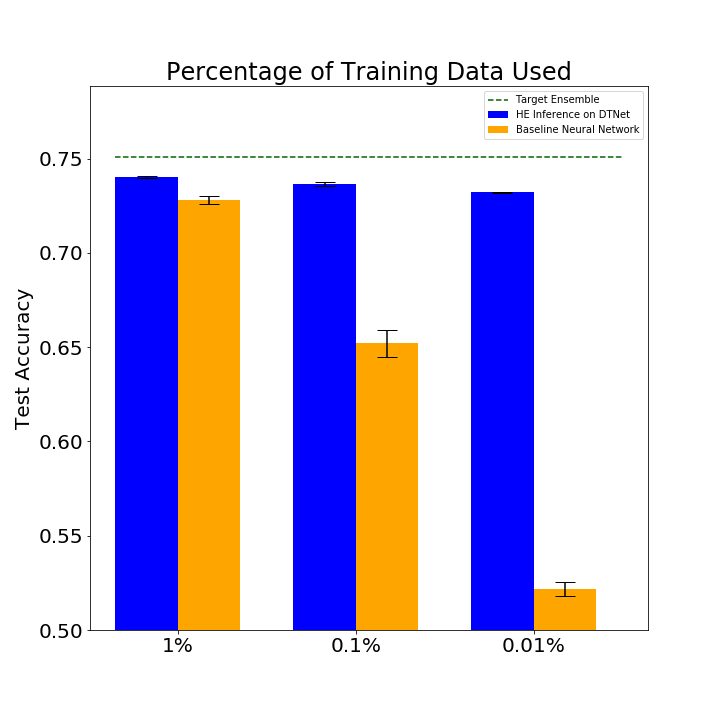}
    \caption{Allstate Dataset}
\end{subfigure}\\
\begin{subfigure}[b]{0.35\linewidth}
    \includegraphics[width=\linewidth]{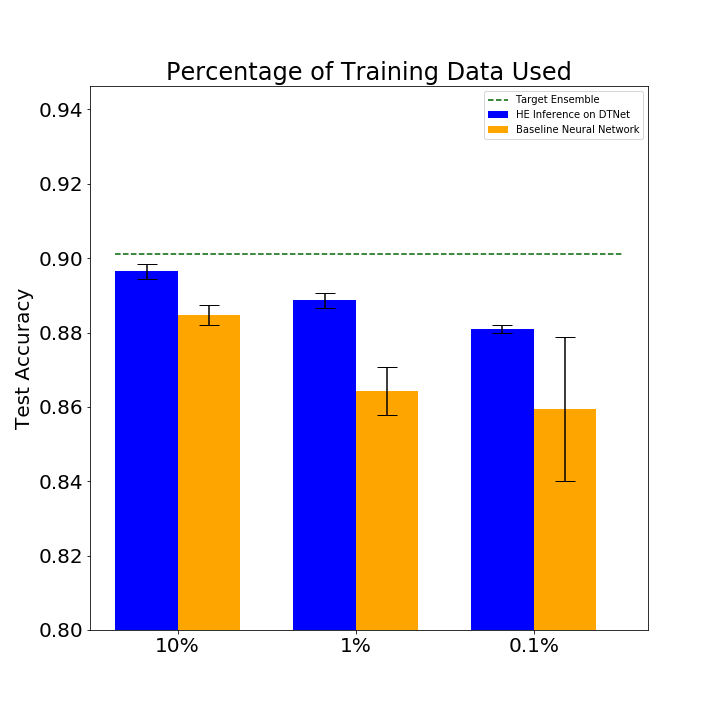}
    \caption{Bank Dataset}
\end{subfigure}
\begin{subfigure}[b]{0.35\linewidth}
    \includegraphics[width=\linewidth]{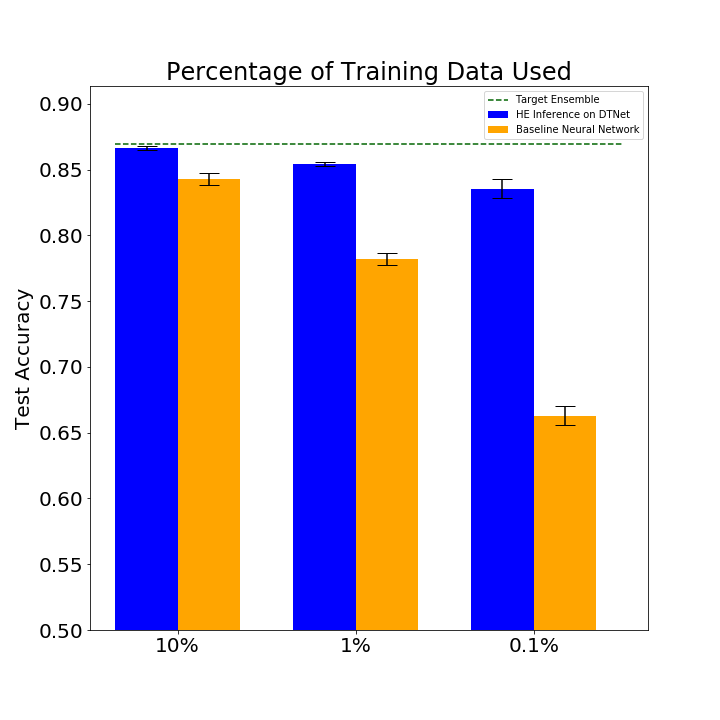}
    \caption{Magic Dataset}
\end{subfigure}

\caption{Comparison of mean test accuracy (and standard deviation) of the DTNets  against the target ensemble (dashed line) and the baseline neural network (NN).  For each dataset, size of the available training data ($\widehat{\mathcal{T}}$) is varied to be in the order of $[10^3, 10^2, 10^1]$ samples. While the DTNet uses only unlabeled samples, the baseline NN uses the same samples along with true labels (and involves no actual transfer of knowledge from the ensemble). DTNet consistently performs significantly better than the baseline NN and in most cases is very close to the target ensemble accuracy.\label{fig:expts}}
\end{figure*}

{\bf Accuracy Results:} We intend to show that the test accuracy of the depth constrained student model (DTNet) on homomorphically encrypted data is close to unencrypted inference accuracy of a given target ensemble. To this end, we perform $36$ experiments spanning across the four datasets and summarize the results in Figure~\ref{fig:expts}. For each dataset and each training datasize, we repeat the following steps $3$ times: (a) train an ensemble on the original data $\mathcal{T}$, (b) obtain the DTNet on using $\mathcal{S}$ and $\mathcal{V}$ (unlabeled data), (c) train the baseline on $\widehat{\mathcal{T}}$ along with the true labels, and (d) compute the mean test accuracy and standard deviation. Figure~\ref{fig:expts} compares the mean test accuracy (and deviation) of DTNets against the target ensemble (dashed line) and the baseline neural network. 

In all the experiments, the DTNets consistently outperform the baseline and in most cases endure an accuracy loss of less than $1-2\%$ compared to the target ensemble. As the amount of training data available decreases, the performance of the baseline drops considerably, but the DTNet sees only a mild dip in accuracy, indicating a clear transfer of knowledge from the ensemble. This also demonstrates the superiority of the decision boundaries learned using boosted decision trees and justifies our choice of decision tree ensembles as the teacher model for such tabular datasets.

\subsection{Efficiency of Homomorphic Inference on DTNets} 

While the DTNet search and its training are performed in the unencrypted space, the inference is performed in the encrypted domain. Our FHE implementation (a) can perform efficient batch processing by exploiting SIMD operations and multi-threading strategies (b) has a strong security level of $134$ bits (recommended minimum is $128$ bits).

{\bf FHE Library and Parameters:}  Our FHE module uses the CKKS scheme implemented using HELib library~\cite{HElib} (commit number \texttt{7427ed3709bb9872835324dd0007a97b3ca3baca}). The parameters of the encryption scheme were chosen to allow sufficient depth for inferencing on encrypted data. More precisely, we chose the depth parameter (nBits) to be $800$ bits and the underlying cyclomatic field of the encryption scheme was set to $m = 2^{16} = 65536$. This gives us a strong security of $134$ bits and we can pack $m/4 = 16384$ floating point numbers in each ciphertext for SIMD operations. Note that we choose a sufficiently large initial scaling factor (modulus) in CKKS such that the final prediction results (class labels) are exactly identical to the plaintext inference results. There may be some differences in the least significant bits of the score, but we did not observe any case where it impacted the final classification result. The only accuracy loss is due to the transformation from decision tree ensemble to depth-constrained neural network.

{\bf Model Parameter Extraction and Packing Strategy:}
The first step in our FHE implementation is to extract the weights in each of the layers of the DTNet and encrypt them. We use Keras's \texttt{get\_weights} function to obtain the weights and biases in each layer of the DTNet. We adopt a {\em vertical repetition} packing strategy to pack our weights into ciphertexts \cite{gilad2016cryptonets}. To this end, given the weights of the network $w_{e}$, for each link $e$ in the DTNet, we construct a ciphertext packing by repeating $w_e$ $C$ times, where $C$ is the number of {\em slots} in the ciphertext packing. Given a batch of test dataset $B$, we first partition it into multiple mini-batches $[B_i]_{i\in |B|/C}$ of capacity $C$. For each mini-batch $B_i$ with points $\mathbf{x}_{ij} = (x^k_{ij})_{k\in [d]}$, we construct $d$ ciphertexts $\mathbf{c}_k = \mathcal{E}([x^k_{i1}, x^k_{i2},\ldots, x^k_{iC}])$. 

{\bf Multi-threading Strategy:} The above packing approach allows us to effectively use the SIMD operations to process $C \approx 16k$ data points in a single inference execution. To further scale the system, we adopt the following multi-threading strategies to speed up the inference executions on multi-core cloud environments. We perform our experiments on a cloud environment where each node is comprised of 2x Power-9 processors clocked at $3.15$ GHz with $512$ GB RAM. Each processor contains 20 cores with 4 hardware threads (160 logical processors per node).

Since neuron computations in each layer are independent, we process each neuron using a separate thread (based on thread availability). To further exploit the availability of threads, we parallelize the dot product evaluations within each node in the network. Suppose that the weights of links incident on the node are $\mathbf{w} = [w_1, w_2, \ldots, w_n]$ and the corresponding inputs are $\mathbf{x} = [x_1, x_2, \ldots, x_n]$. The dot product $\mathbf{w}\cdot \mathbf{x} = \sum_{i\in [n]}w_ix_i$ can be computed recursively by partitioning $\mathbf{w}$ into roughly equal parts $\mathbf{w}_1, \mathbf{w}_2$ and $\mathbf{x}$ into $\mathbf{x}_1$, $\mathbf{x}_2$. Then, $\mathbf{w}\cdot \mathbf{x} = \mathbf{w}_1\cdot \mathbf{x}_1 + \mathbf{w}_2\cdot \mathbf{x}_2$. The computations in each level of the corresponding binary recursion tree can be performed on a different thread. However, we observed that parallelization of the recursive dot product beyond a certain depth is not beneficial because context switching between threads and locking time tend to dominate.

\begin{figure}[h]
\centering
\includegraphics[width=\linewidth]{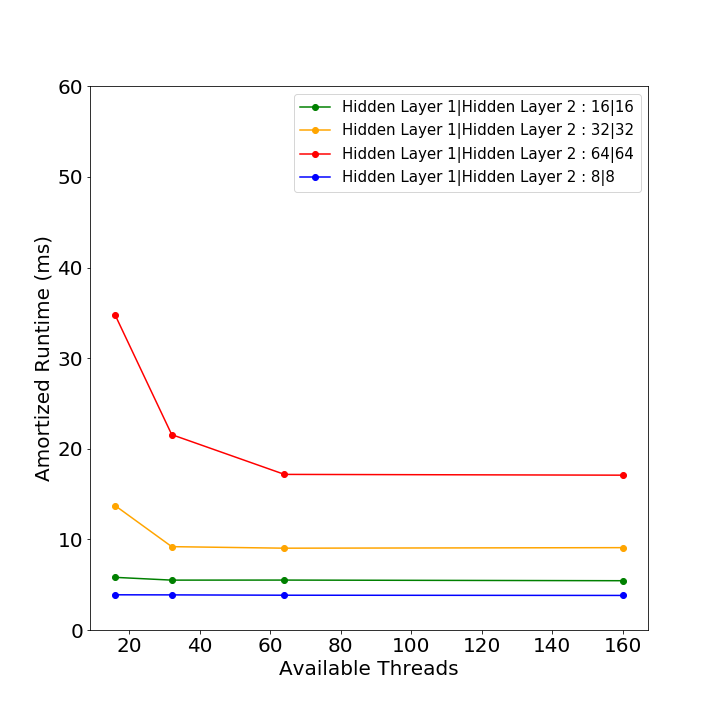}
\caption{Amortized execution time based on the proposed multithreading strategy for encrypted batched inference over $200,000$ test datapoints.}
\label{fig:threading}
\end{figure}

{\bf Efficiency Results:} Figure~\ref{fig:threading} shows the amortized execution times based on the above multi-threading approach for different architectures of the DTNets on $200,000$ points (Allstate). Even for the larger $64|64$ network, the proposed approach allows us to process the above $200,000$ points in little less than an hour\footnote{Note that we discount the initial encryption setup and preprocessing times, which is a one time cost.} on a compute node with $64$ logical threads. This is equivalent to an amortized time of around $18$ ms per query. Since each inference batch packs $C \approx 16k$ queries, the latency of prediction is approximately $295$ seconds.

It must be highlighted that the ciphertext packing strategy used in this work has high latency for a single inference and high memory requirements. This could be addressed using different packing strategies as discussed in \cite{brutzkus2019low}, which can reduce the latency at the cost of decreasing the throughput. Furthermore, the literature on encrypted inference use (i) different datasets (e.g., MNIST, CIFAR-10), (ii) different encryption schemes (e.g., BFV, TFHE), (iii) different libraries (e.g., SEAL, Helib), and (iv) different platforms (e.g., x86, Power). Hence, it is
not possible to directly compare the execution times reported in this work with those reported in the literature. Even for the same FHE scheme, say CKKS, the choice of library and implementation platform have a significant impact on the execution times due to differences in various low-level (library and platform) optimizations. Therefore, direct comparison with other reported implementations is not possible.

{\bf Efficiency Comparison with Soft Comparator Approach:} As mentioned earlier, there is no prior work that we are aware of that addresses the FHE inference on ensemble decision trees satisfactorily. Even in the case of a single decision tree, the prior works fail to report either accuracy loss or the runtime guarantees. Hence, we estimate the efficiency gain compared to the standard practice of replacing the hard comparisons in a decision tree with high depth soft comparators~\cite{cheon2019numerical}. Assuming a perfect multi-threaded implementation on a machine with 32 logical threads, we need $4$ sequential evaluations of decision trees to process an ensemble of 100 decision trees. Further, assuming a vertical packing of the ciphertexts and a tree of depth $5$, we must process $31$ comparison operations in sequence (note that all the threads have been used up for processing the decision trees in parallel). Assuming $5-10$ bootstrapping operations per comparison operation, each taking an amortized runtime of $0.01$ second~\cite{chen2019improved}, we need an amortized runtime of $31\cdot [5-10]\cdot 0.01\cdot 4 = 6-12$ seconds per point. This is a very conservative estimate and considers only the comparison operations (note that there aggregation of the results from various trees can be highly non-trivial as well in the encrypted space). Since the amortized runtime of our DTNet approach is in the order of tens of milliseconds, we achieve roughly three orders of magnitude improvement over the standard approach. 

\section {Conclusions and future work} 
We proposed the first systematic approach to perform multiplicative depth constrained knowledge distillation that enables efficient encrypted inference. We demonstrated the power of this approach by  performing highly efficient encrypted inference on complex ensembles of decision trees. We further demonstrated the high scalability of our system in multi-core cloud environments. On the theoretical side, a more refined architectural search needs to developed in the future. On the practical side, reducing the inference latency and the memory requirements using alternative ciphertext packing strategies as well as various code optimizations needs to be explored.

\bibliographystyle{abbrv}
\bibliography{references}

\end{document}